%% file: main.tex
\documentclass[aip, amsmath, amssymb,reprint]{revtex4-1}

\usepackage[utf8]{inputenc}
\usepackage[T1]{fontenc}
\usepackage{mathptmx}
\usepackage{etoolbox}

\usepackage{graphicx}
\usepackage{amsmath}
\usepackage{color}
\usepackage{comment}
\usepackage{hyperref}
\usepackage{multirow}
\usepackage{subfig}
\usepackage{braket}
\usepackage{booktabs}
\usepackage{siunitx}
\usepackage[version=2]{mhchem}
\usepackage[normalem]{ulem}
\usepackage{bm}

\makeatletter
\def\@email#1#2{%
 \endgroup
 \patchcmd{\titleblock@produce}
  {\frontmatter@RRAPformat}
  {\frontmatter@RRAPformat{\produce@RRAP{*#1\href{mailto:#2}{#2}}}\frontmatter@RRAPformat}
  {}{}
}%
\makeatother

\definecolor{ao}{rgb}{0.0, 0.5, 0.0}

\newcommand*{\citen}[1]{%
  \begingroup
    \romannumeral-`\x % remove space at the beginning of \setcitestyle
    \setcitestyle{numbers}%
    \cite{#1}%
  \endgroup   
}

\begin{document}

\title{Polaritonic Chemistry using the Density Matrix Renormalization Group Method}

\author{Mikuláš Matoušek}
\email{mikulda@volny.cz}
\affiliation{J. Heyrovsk\'{y} Institute of Physical Chemistry, Academy of Sciences of the Czech \mbox{Republic, v.v.i.}, Dolej\v{s}kova 3, 18223 Prague 8, Czech Republic}
\affiliation{Faculty of Mathematics and Physics, Charles University, Prague, Czech Republic}

\author{Nam Vu}
\affiliation{Department of Chemistry, University of North Carolina Charlotte, Charlotte, North Carolina 28223, United States}

\author{Niranjan Govind}
\affiliation{Physical and Computational Sciences Directorate, Pacific  Northwest National Laboratory, Richland, Washington 99352, United States}
\affiliation{Department of Chemistry, University of Washington, Seattle, Washington 98195, United States}

\author{Jonathan J. Foley IV}
\email{jfoley19@charlotte.edu}
\affiliation{Department of Chemistry, University of North Carolina Charlotte, Charlotte, North Carolina 28223, United States}

\author{Libor Veis}
\email{libor.veis@jh-inst.cas.cz}
\affiliation{J. Heyrovsk\'{y} Institute of Physical Chemistry, Academy of Sciences of the Czech \mbox{Republic, v.v.i.}, Dolej\v{s}kova 3, 18223 Prague 8, Czech Republic}


\begin{abstract}
The emerging field of polaritonic chemistry explores the behavior of 
%organic 
molecules under strong coupling with cavity modes.
Despite recent developments in {\it ab initio} polaritonic methods for simulating polaritonic chemistry under electronic strong coupling, their capabilities are limited, especially in cases where the molecule also features strong electronic correlation.
To bridge this gap, we have developed a novel method for cavity QED calculations utilizing the Density Matrix Renormalization Group (DMRG) algorithm in conjunction with the Pauli-Fierz Hamiltonian.
Our approach is applied to investigate the effect of the cavity on the \mbox{S$_0$-S$_1$} transition of $n$-oligoacenes, with $n$ ranging from 2 to 5, encompassing 22 fully correlated $\pi$ orbitals in the largest pentacene molecule. Our findings indicate that the influence of the cavity intensifies with larger acenes. Additionally, we demonstrate that, unlike the full determinantal representation, DMRG efficiently optimizes and eliminates excess photonic degrees of freedom, resulting in an asymptotically constant computational cost as the photonic basis increases.
\end{abstract}

\maketitle

\section{Introduction}
\label{section_introduction}
\input{introduction}

In what follows, we briefly review the basics of the DMRG method and outline the extensions necessary for QED-DMRG. Since our DMRG implementation (MOLMPS program \cite{Brabec2020}) employs the renormalized operators rather than the matrix product operators (MPOs), the presentation follows the original renormalization group picture \cite{DMRG_White}. In order to demonstrate the accuracy of QED-DMRG, we present its application on the eigenstates of the Pauli-Fierz (PF) Hamiltonian tuned to the excitation from the ground to the first excited singlet state (S$_0$ to S$_1$) of $n-$oligoacenes with $n \in \langle 2, 5 \rangle$. For a discussion on the excited states of $n-$oligoacenes using a broad spectrum of excited-state theoretical approaches, we refer the reader to Ref. \citenum{lopata:2011} and references therein.

\section{Theory}
\label{section_theory}

\subsection{Quantum chemical DMRG method}

While the Density Matrix Renormalization Group (DMRG) method has origins in solid state physics\cite{DMRG_White, White-1993}, it is already established also in the realm of quantum chemistry\cite{chan_review, Szalay2015, reiher_perspective}, sometimes abbreviated as QC-DMRG. Here the DMRG algorithm is used to find the eigenstates of the electronic Hamiltonian
\begin{equation}
    H_\mathrm{el.} = \sum_{\sigma} \sum_{pq} h_{pq} a_{p_{\sigma}}^{\dagger} a_{q_{\sigma}} +
    \frac{1}{2} \sum_{\sigma \sigma^{\prime}}\sum_{pqrs} \langle pq | rs \rangle a_{p_{\sigma}}^{\dagger} a_{q_{\sigma^{\prime}}}^{\dagger} a_{s_{\sigma^{\prime}}} a_{r_{\sigma}},
  \label{ham_sec_quant}
\end{equation}

\noindent
where $h_{ij}$ and $\langle ij | kl \rangle$ denote standard one and two-electron integrals in the restricted molecular orbital (MO) basis, and $\sigma$ and $\sigma^{\prime}$ denote spin, $\sigma, \sigma^{\prime} \in \{ \uparrow, \downarrow \}$.

The DMRG method is based on the Matrix Product State (MPS) \cite{Schollwock2011} wave function ansatz, in which the exact FCI wave function (in the occupation basis representation) 

\begin{equation}
    |\Psi_\mathrm{FCI}\rangle = \sum_{n_1 \cdots n_k} c^{n_1 n_2 \cdots n_k} |n_1 n_2 \cdots n_k \rangle
    \label{wf_fci}
\end{equation}

\noindent
is factorized into a linear tensor network

\begin{equation}
    |\Psi_\mathrm{MPS}\rangle = \sum_{n_1 \cdots n_k} \sum_{i_1 \cdots i_{k-1}} A[1]^{n_1}_{i_1} A[2]^{n_2}_{i_1 i_2} \cdots A[k]^{n_k}_{i_{k-1}} |n_1 \cdots n_k \rangle.
\end{equation}

\noindent
The tensors $A[j]$ have three indices. One physical, here labeled $n_j$, corresponding to a physical degree of freedom. In quantum chemistry these are the possible occupations of a single molecular orbital ( $|\;\;\;\;\rangle,|\!\!\uparrow\;\;\rangle,|\;\downarrow\rangle,|\!\!\uparrow\downarrow\rangle$ ), leading to a dimension of four.
The other two ($i_{j-1},i_j$) are called virtual, and they are contracted with the neighboring tensors. The dimension of these indices, called the bond dimension,
%can be chosen arbitrarily
is the main parameter controlling both the accuracy and the computational cost \cite{Szalay2015}. Since the bond dimension is in practical approximate calculations bounded, the number of DMRG variational parameters is, in contrast to FCI, polynomial. 

The optimization of the individual tensors in the network is then performed sequentially in a process called sweeping. 
The standard two-site algorithm is based on the separation of the tensor network into four parts, the two sites taken explicitly ($j$, $j+1$), all the sites to the left of the explicit sites ($i<j$, so called left block) and the 
sites on the right ($i>j+1$, right block). It provides the wave function in the two-site MPS form \cite{Schollwock2011}

\begin{small}
\begin{equation}
  \ket{\Psi_{\text{2s-MPS}}} = \sum_{\{n\}} \sum_{\{i\}} A^{n_1}_{i_1} \cdots A^{n_{j-1}}_{i_{j-2} i_{j-1}}  \Psi^{n_j n_{j+1}}_{i_{j-1} i_{j+1}} A^{n_{j+2}}_{i_{j+1} i_{j+2}} \cdots A^{n_k}_{i_{k-1}} |n_1 \cdots n_k \rangle,
  \label{2smps}
\end{equation}
\end{small}

\noindent
where we have for simplicity omitted the MPS tensor square brackets with the site indices. The uncontracted virtual indices on the left and right blocks form the orthonormal many-particle bases ($\{\ket{l}\}, \{\ket{r}\}$), which span a subspace of the full Hilbert space of a given block. The orthonormality stems from the fact that the MPS tensors formed during the DMRG sweeping via the singluar value decomposition (SVD, see below) fulfil the following conditions

\begin{eqnarray}
\sum_{n_j}(\mathbf{A}[j]^{n_j})^{\dagger} \mathbf{A}[j]^{n_j} & = & \mathbf{I} \qquad j \in \text{left} \\
\sum_{n_j}\mathbf{A}[j]^{n_j} (\mathbf{A}[j]^{n_j})^{\dagger} & = & \mathbf{I} \qquad j \in \text{right}
\end{eqnarray}

The four-index tensor $\Psi^{n_j n_{j+1}}_{i_{j-1} i_{j+1}}$ in Eq. \ref{2smps} is
in each iteration of the sweeping procedure obtained by solving the Schr\"{o}dinger equation projected onto the product space of the left block, two explicit sites, and the right block. The effective equation has, due to the orthonormality of the left and right block bases mentioned above, a form of a standard eigenvalue problem

\begin{equation}
  \label{h_mat}
  H_{\text{el.}} \ket{\Psi} = E \ket{\Psi},
\end{equation}

\noindent
which is solved by means of iterative solvers such as the Davidson algorithm \cite{Davidson_1975}. The wave function is expanded as

\begin{equation}
  \label{wf}
  \ket{\Psi} = \sum_{l n_j n_{j+1} r} \Psi^{n_j n_{j+1}}_{l r} \ket{l} \otimes \ket{n_j} \otimes \ket{n_{j+1}} \otimes \ket{r}.
\end{equation}

\noindent
Notice that $\ket{l}$ labels basis states of the left block, which contains $j-1$ sites, i.e. they correspond to the aforementioned uncontracted MPS virtual index $i_{j-1}$.

\begin{small}
\begin{equation}
    \ket{l} \equiv |\psi_{i_{j-1}}\rangle = \sum_{n_1 \cdots n_{j-1}} \sum_{i_1 \cdots i_{j-2}} A^{n_1}_{i_1} A^{n_2}_{i_1 i_2} \cdots A^{n_{j-1}}_{i_{j-2} i_{j-1}} |n_1 \cdots n_{j-1} \rangle.
\end{equation}
\end{small}

\noindent
Similarly $\ket{r}$ corresponds to the uncontracted virtual index $i_{j+1}$.
%The electronic Hamiltonian (\ref{ham_sec_quant}) is thus expressed in the basis of a tensor product between individual spaces and diagonalized.
%The trick to obtaining a polynomial scaling is to not work with the explicit representation of the states, but instead keep the explicit form of the different combinations of creation and annihilation operators acting on the left and right blocks. 

In DMRG, the explicit (determinant) representations of the complicated many-particle bases ($\{\ket{l}\}$, $\{\ket{r}\}$) are not stored, because it would lead to the original exponential scaling. Instead, the matrix representations of second-quantized operators needed for the action of the Hamiltonian on a wave function (see Eq. \ref{h_mat}) are formed and stored.
The matrix representations of a single site annihilation operators (in the \mbox{$\{|\;\;\;\;\rangle, |\!\!\uparrow\;\;\rangle,|\;\downarrow\rangle,|\!\!\uparrow\downarrow\rangle\}$} basis) read

\begin{equation}
  \label{creation_ops}
  a_{\uparrow} = \begin{pmatrix} 0 & \phantom{-}0 & \phantom{-}1 & \phantom{-}0 \\ 0 & \phantom{-}0 & \phantom{-}0 & \phantom{-}1 \\ 0 & \phantom{-}0 & \phantom{-}0 & \phantom{-}0 \\ 0 & \phantom{-}0 & \phantom{-}0 & \phantom{-}0 \end{pmatrix}, \quad a_{\downarrow} = \begin{pmatrix} 0 & \phantom{-}1 & \phantom{-}0 & \phantom{-}0 \\ 0 & \phantom{-}0 & \phantom{-}0 & \phantom{-}0 \\ 0 & \phantom{-}0 & \phantom{-}0 & -1 \\ 0 & \phantom{-}0 & \phantom{-}0 & \phantom{-}0 \end{pmatrix} .
\end{equation}

When working with the enlarged left block (L) as a tensor product of the left block states with the states on the explicit left site and similarly the right enlarged block (R), the Hamiltonian for this bipartite splitting reads

\begin{equation}
  H_{\text{el.}} = H_\mathrm{L} \otimes I_\mathrm{R} + I_\mathrm{L} \otimes H_\mathrm{R} + \sum_{\alpha} H_\mathrm{L}^{\alpha} \otimes H_\mathrm{R}^{\alpha},
\end{equation}

\noindent
where $H_{\mathrm{L}/\mathrm{R}}$ represent the enlarged left/right block Hamiltonians, i.e. all indices in Eq. \ref{ham_sec_quant} belonging to the corresponding blocks and the last summation term represents the interaction between enlarged left and right blocks with indices split between both blocks. Let us demonstrate the main strategies on the example of the simpler one-electron Hamiltonian

\begin{equation}
  H_{\text{one-el.}} = \sum_{pq} h_{pq} a^{\dagger}_{p\uparrow} a_{a\uparrow} + \sum_{pq} h_{pq} a^{\dagger}_{p\downarrow} a_{a\downarrow}.
\end{equation}

\noindent
Here, the interaction between the enlarged blocks consists of the two contributions

\begin{eqnarray}
H_{\text{one-el.}}^{\text{int}} & = & \sum_{q\in \mathrm{R}}
    \left\lbrace 
    \sum_{p\in \mathrm{L}} h_{pq} a_{p\uparrow}^\dagger
\right\rbrace_\mathrm{L}\otimes
(a_{q\uparrow})_\mathrm{R} \nonumber \\
 & + & \sum_{q\in \mathrm{R}}
\left\lbrace 
    \sum_{p\in \mathrm{L}} h_{pq} a_{p\downarrow}^\dagger
\right\rbrace_\mathrm{L}\otimes
(a_{q\downarrow})_\mathrm{R} \nonumber \\
& - &  \mathrm{h.\,c.}
\label{h_int_oneel}
\end{eqnarray}

In order to reduce the number of matrix-matrix multiplications during the action of the Hamiltonian on a wave function, which are the most CPU-demanding tasks, the efficient QC-DMRG codes work with the so called presummed (or partially summed) operators \cite{Xiang-1996}, i.e. intermediates formed by contraction of operator matrices with MO integrals. In case of the one-electron Hamiltonian, Eq. \ref{h_int_oneel}, the enlarged left block one-electron presummed operators are encapsulated in the curly brackets and are defined as

\begin{eqnarray}
  A^{\uparrow}_q & = & \sum_{p\in \mathrm{L}} h_{pq} a_{p\uparrow}^\dagger, \qquad q \notin L \label{opAup}\\
  A^{\downarrow}_q & = & \sum_{p\in \mathrm{L}} h_{pq} a_{p\downarrow}^\dagger, \qquad q \notin L \label{opAdown}
\end{eqnarray}

The interaction terms are always composed of presummed operators on the one block and normal operators on the other block. In the case of one-electron Hamiltonian (\ref{h_int_oneel})

\begin{equation}
  H_{\text{one-el.}}^{\text{int}} = \sum_{q \in R} (A^{\uparrow}_q)_\text{L} \otimes (a_{q\uparrow})_\text{R} + \sum_{q \in R} (A^{\downarrow}_q)_\text{L} \otimes (a_{q\downarrow})_\text{R} - \text{h.c.}
\end{equation}

\noindent
Since the presummed operators are for efficiency reasons formed on the longer block, a switching between normal and presummed operators has to be done in the middle of each sweep.

\begin{comment}  
The full Hamiltonian then needs to be split into contributions from operators acting on the different subspaces.
If we define the enlarged left block $\mathrm{L}$ as a tensor product of the left block states with the states on the left site, and similarly the right enlarged block $\mathrm{R}$, we can split the Hamiltonian into those subspaces. For example the one electron part $\sum_{i,j} h_{ij} a^\dagger_i a_j$ gives these contributions:

\begin{align*}
    \left\lbrace 
\sum_{i,j\in \mathrm{L}}  h_{ij} a_i^\dagger a_j 
\right\rbrace_\mathrm{L}&\otimes\left\lbrace
\mathrm{id} 
\right\rbrace_\mathrm{R},\\
    \sum_{j\in \mathrm{R}}
    \left\lbrace 
    \sum_{i\in \mathrm{L}} h_{ij} a_i^\dagger
\right\rbrace_\mathrm{L}&\otimes\left\lbrace
\mathrm{a_j}
\right\rbrace_\mathrm{R}  - \mathrm{h.\,c.},\\
    \left\lbrace 
    \mathrm{id}
\right\rbrace_\mathrm{L}&\otimes\left\lbrace
 \sum_{i,j\in \mathrm{R}} h_{ij} \mathrm{a_i^\dagger a_j}
\right\rbrace_\mathrm{R},\\
\end{align*}
where each of the operators in the enlarged left and right blocks has to be further written as a combination of products of operators on the respective smaller block and on the added site.
\end{comment}

The transition from one iteration of the sweep to another is carried out by means of the renormalization procedure. 
In this step, the basis of one of the blocks is, in the direction of the sweep, enlarged by one new site and the operators needed for the action of the Hamiltonian on the trial wave function are transformed into the new basis. The complementary block is, on the other hand, reduced by one site.
This effectively moves us by one tensor in the MPS and the diagonalization of the effective Hamiltonian (Eq.~\ref{h_mat}) can be repeated.
The exact basis obtained by enlarging by one site would be a tensor product between the original basis and the basis of the new site. 
This is unacceptable, as the size of the new basis would be a product of the sizes of the old basis and the basis of the added site. 
The choice of truncation of the new basis gave the name to the method, the basis is truncated so that the density matrix of the new block is changed as little as possible.
This is done by diagonalizing the density matrix and keeping only the largest eigenvalues. 
When the wave function expansion coefficients $\Psi_{lr}^{n_j n_{j+1}}$ (\ref{wf}) are reshaped into the matrix form $\psi_{(l n_j), (n_{j+1} r)}$, the aforementioned reduced density matrices can be computed in the following way

\begin{eqnarray}
  \bm{\rho}^L & = & \bm{\psi} \bm{\psi}^{\dagger}, \\
  \bm{\rho}^R & = & \bm{\psi}^{\dagger} \bm{\psi}.
\end{eqnarray}

\noindent
For the transition to the next iteration, all operator matrices formed for the enlarged block have to be renormalized

\begin{equation}
  \label{renorm_mat}
  \bm{A}^{\prime} = (\bm{O}^L)^\dagger \bm{A} \bm{O}^L ,
\end{equation}

\noindent
where $\bm{A}$ represents an operator matrix in the non-truncated ($4M$-dimensional) basis and $\bm{A}^{\prime}$ is the renormalized matrix representation in the truncated ($M$-dimensional) basis.

The procedure outlined above effectively performs an SVD decomposition on the wave function (\ref{wf})

\begin{equation}
  \Psi^{n_j n_{j+1}}_{i_{j-1} i_{j+1}} \xrightarrow[]{\text{SVD}} A_{i_{j-1} i_j}^{n_j} M^{n_{+1}}_{i_j i_{j+1}},
  \label{svd}
\end{equation}

\noindent
where the new MPS tensor $A[j]$ is formed and the three-leg tenosr $M[j+1]$ represents the rest of the SVD factorization. SVD in fact 
produces the best approximation of bipartite wave functions \cite{Schollwock2011}. 
The sum of the discarded eigenvalues is then called the Truncation Error (TRE).

\subsection{Cavity QED-DMRG method}
\label{subsec_qed_dmrg}

The generalization of DMRG to multiple particle types is straightforward.
The Hamiltonian now contains a mix of creation and annihilation operators corresponding to the different particle types, which separately satisfy own commutation/anti-commutation rules.
Each of the particle types is assigned its own set of sites in the MPS wave function and renormalized operators for all particle types, including the mixed terms decribing the interaction between models, have to be formed. Previously, the QC-DMRG method has been generalized for the beyond-Born-Oppenhemier nuclear-electronic all-particle (NEAP) calculations \cite{Muolo2020, Feldmann2022} and applied on the proton-electron problems, i.e. spin-$\frac{1}{2}$ fermionic interactions. Herein, we generalize the QC-DMRG method for applications in polaritonic chemistry in which electronic (fermionic) and photonic (bosonic) degrees of freedom are treated on equal footing. Below, we give the details specific to this generalization.

%Cavity-QED calculations are usually based on the Pauli-Fierz Hamiltonian
We will restrict ourselves to the Pauli-Fierz (PF) Hamiltonian within the dipole approximation, which describes coupling of molecular systems to a single cavity (photonic) mode
\begin{comment}
\begin{equation}
    \hat{H}_\mathrm{PF} =  \hat{H}_\mathrm{QC} + \omega b^\dagger b -\sqrt{\frac{\omega}{2}}\mathbf{\lambda}\cdot\left(\mathbf{\hat{d}} - \left\langle \mathbf{d_e} \right\rangle \right)\left( b + b^\dagger \right) + \left(\lambda\cdot\left(\mathbf{\hat{d}} - \left\langle \mathbf{d_e} \right\rangle \right)\right)^2,
\end{equation}
\end{comment}

\begin{comment}
\begin{equation}
    \hat{H}_\mathrm{PF} =  \hat{H}_\mathrm{el} + \omega b^\dagger b -\sqrt{\frac{\omega}{2}}\left(\hat{d} + \left\langle d_\lambda \right\rangle \right)\left( b + b^\dagger \right) + \frac{1}{2}\left(\hat{d} + \left\langle d_\lambda \right\rangle \right)^2,
\end{equation}
\end{comment}

\begin{equation}
  H_{\text{PF}} = H_{\text{el.}} + \omega b^\dagger b -\sqrt{\frac{\omega}{2}} \bm{\lambda} \cdot \bm{\mu} \left( b^\dagger + b \right) + \frac{1}{2} \left( \bm{\lambda} \cdot \bm{\mu} \right)^2,
  \label{ham_pf}
\end{equation}

\noindent
where $H_\mathrm{el.}$ is the electronic Hamiltonian (\ref{ham_sec_quant}), $\omega$ is the cavity photon frequency, $\bm{\mu}$ represents the molecular dipole operator, $\bm{\lambda}$ is a coupling vector, and $b^{\dagger}$, $b$ denote photonic creation and annihilation operators. The second term in Eq. \ref{ham_pf} corresponds to the harmonic oscillator Hamiltonian of the bare cavity mode, the third term represents the so called bilinear coupling, and the last one is the dipole self-energy term. We will assume the Cartesian coordinate system, thus $\bm{\lambda}$ and $\bm{\mu}$ have $x$, $y$, and $z$ components. It is convenient to define the scalar molecular dipole coupling operator as $d = \bm{\lambda} \cdot \bm{\mu}$.

Following the previous work on QED-CASCI \cite{QED-CASCI}, we compare two approaches differing by the input molecular orbitals. The simpler scenario corresponds to using the standard Hartree-Fock (HF) orbitals and searching for the eigenstates of the PF Hamiltonian presented in Eq. \ref{ham_pf}. In this case and similarly to Eq. \ref{wf_fci}, the QED-FCI wave function is expanded in the so called particle number (PN) basis

\begin{equation}
    |\Psi_\mathrm{QED-FCI}\rangle = \sum_{N} \sum_{n_1 \cdots n_k} c^{N n_1 n_2 \cdots n_k} \ket{N^{\text{ph.}}} \otimes |n_1 n_2 \cdots n_k ^{\text{el.}}\rangle,
    \label{wf_qed_fci}
\end{equation}

\noindent
where $\ket{N^{\text{ph.}}}$ denotes the photonic PN basis, which comprises states $\ket{0}, \ket{1}, \ket{2}, \dots \ket{n_\text{max}}$, $n_\text{max}$ being the maximum photon occupation.

The second strategy is to perform the QED-HF calculation
and transform from the PN basis into the coherent state (CS) basis \cite{klauder:1968, Foley_CPR_2023} by means of the unitary transformation

\begin{equation}
U_{\text{CS}}=\exp{(z(b^\dagger-b))}, \qquad z = \frac{-\left\langle \bm{\mu}_{\text{QED-HF}} \right\rangle\cdot\bm{\lambda}}{\sqrt{2\omega}}.
\end{equation}

\noindent
In order to keep the same wave function expansion as in Eq. \ref{wf_qed_fci}, $U_{\text{CS}}$
is applied at the level of the Hamiltonian (\ref{ham_pf}), which yields the PF Hamiltonian in the coherent state basis 

\begin{equation}
    H_\mathrm{CS} =  H_\mathrm{el.} + \omega b^\dagger b -\sqrt{\frac{\omega}{2}}\left(d_{\text{e}} - \left\langle d_\text{e} \right\rangle \right)\left( b + b^\dagger \right) + \frac{1}{2}\left(d_{\text{e}} - \left\langle d_\text{e} \right\rangle \right)^2,
    \label{ham_cs}
\end{equation}

\noindent
%\sout{where  $d_{\text{e}}$ denotes the electronic part of the dipole coupling operator, which can be written in the MO basis as }
where denoted as $d_{\text{e}}$ and $ \left\langle d_\text{e} \right\rangle$ we keep only the electronic part of the dipole coupling operator $d$ and of the coupling expectation value $\left\langle d \right\rangle = \left\langle \bm{\mu}_{\text{QED-HF}} \right\rangle\cdot\bm{\lambda}$, because the nuclear parts cancel under the Born-Oppenheimer approximation.

We would like to point out that the reference state $\ket{0^{\text{ph.}}} \otimes \ket{\phi_0^{\text{el.}}}$ transformed into the CS basis formally includes an infinite number of photon occupation states, due to the exponential form of $U_{\text{CS}}$ \cite{QED-CASCI}. Consequently, it was shown to outperform the PN basis in the correlated QED-CASCI calculations \cite{QED-CASCI}.

\subsection{Implementation details}

\begin{comment}
\begin{equation}
q_{\text{e}}=\sum_{i,j} q_{ij} a^\dagger_{i\uparrow} a_{j\uparrow} + q_{ij} a^\dagger_{i\downarrow} a_{j\downarrow},
\end{equation}

\noindent
with the quadrupole coupling integral \mbox{$q_{ij} = -\int \phi^*_{i}(\mathbf{r})\,(\bm{\lambda}\cdot\mathbf{r})^2\,\phi_{j}(\mathbf{r})\,\mathrm{d}\mathbf{r}$}. 
\end{comment}

\begin{comment}
Finally the selection of the background dipole coupling $\left\langle d_\lambda \right\rangle$ depends on the exact choice photonic basis. If we work in the basis with of states with a given number of photons, the background dipole coupling contains just the dipole moment of the nuclei $\left\langle d_\lambda \right\rangle = \left\langle \mathbf{\mu}_\mathrm{nucl} \right\rangle \cdot \lambda$. This formulation has however one caveat, in a truncated photonic basis the energy is not invariant with respect to the choice of coordinate origin for charged molecules.

This can be alleviated by transforming into the coherent state basis, 
using the transformation operator $U=e^{z(b^\dagger-b)}$ with $z = \frac{-\left\langle \mu_{HF} \right\rangle\cdot\lambda}{\sqrt{2\omega}}$, where $\left\langle \mu_{HF} \right\rangle$ is the expectation value of the molecular  dipole moment from a QED-HF calculation. This transformation preserves the structure of the PF Hamiltonian, except for a different background dipole coupling $\left\langle d_\lambda \right\rangle = -\left\langle \mathbf{\mu}_\mathrm{HF} \right\rangle \cdot \lambda$.
\end{comment}

In adapting QC-DMRG for the PF Hamiltonian, the electronic part is identical to QC-DMRG, but we include a single photonic site. Herein, we have restricted ourselves to a single cavity mode, but the Hamiltonian (\ref{ham_pf}), can be easily generalized for multiple cavity modes \cite{Foley_154103} and generalization of the implementation described below is straightforward. Note that the dimension of the photonic site is not 4 as in the case of the electronic sites, but arbitrary, corresponding to the maximum number of photons included in the wave function. 

In our cavity QED-DMRG implementation, we place the photonic site as the leftmost site in the chain.
%The implementation can be simplified using the fact that we have a single photonic site, by placing it as the leftmost site in the chain.
The advantage is twofold. First, at any point over the sweep, the photonic operators are only in the left block.
As a result, the rules for the combination of the operators on different blocks to form the Hamiltonian remain unchanged along the sweep. 
The second advantage is that the right block operators are identical to the QC-DMRG case and we can use the standard CI-DEAS warm-up procedure\cite{LegezaCIDEAS} for the generation of electronic operators in the first sweep. We would like to note that a similar idea of using a single higher dimensional electronic site was used previously \cite{Chan_DMRG_Large_Sites,Legeza_DMRG_large_sites} albeit in a different context. 

%was used in other works using DMRG and placing a single large site to the end of the MPS chain.\cite{Chan_DMRG_Large_Sites,Legeza_DMRG_large_sites}

The single-site matrix representation of the bosonic annihilation operator $b$ in the basis
$\left\lbrace \ket{0}, \ket{1}, \ket{2}, \dots \ket{n_{\text{max}}}\right\rbrace$, analogously to Eq. \ref{creation_ops}, has the matrix representation 
\begin{equation}
b=
    \begin{pmatrix}
        0&\sqrt{1}&0&0\\
        0&0&\sqrt{2}&0\\
        0&0&0&\sqrt{3}\\
        0&0&0&0&\sqrt{4}&&\\        
        &&&&\ddots&\ddots\\
        &&&&&0&\sqrt{n_{\text{max}}}\\
        &&&&&0&0\\
    \end{pmatrix}
\end{equation}
and the creation operator $b^\dagger$ is its hermitian cojugate.

To derive the working form of the PF Hamiltonian, we group the terms containing creation and annihilation operators of only electrons, electrons and photons, and only photons. Here, we present this approach for the CS basis formulation (\ref{ham_cs}). The corresponding PN basis formulation can be readily obtained with a few substitutions in the integrals. 

The $H_{\text{CS}}$ Hamiltonian can be rewritten as

\begin{eqnarray}
    H_\mathrm{CS} & = &  H_\mathrm{el.}^{\prime} -\sqrt{\frac{\omega}{2}} \sum_{\sigma}\sum_{pq}d_{pq}a^{\dagger}_{p\sigma} a_{q\sigma} \left( b + b^\dagger \right) \nonumber \\
    & + & \omega b^\dagger b + \sqrt{\frac{\omega}{2}} \langle d_e \rangle \left( b + b^\dagger \right) + \frac{1}{2} \langle d_e \rangle^2,
    \label{ham_cs_fact}
\end{eqnarray}

\noindent
where the modified electronic Hamiltonain has the same two-body structure as in Eq. \ref{ham_sec_quant}, but with the following MO integrals

\begin{align}
    {\langle p q | r s \rangle}^{\prime} &= \langle p q | r s \rangle + d_{pr}\,d_{qs}, \\
    {h_{pq}}^{\prime} &
     =h_{pq}- \langle d_e \rangle d_{pq} - \frac{1}{2}  q_{pq}
\end{align}

\noindent
and $d_{pq}$ and $q_{pq}$ represent modified electric dipole and electric
quadrupole integrals given by

\begin{eqnarray}
 d_{pq} & = & - \sum_{a \in \{x,y,z\}} \lambda_a \int \phi^*_{p}(\mathbf{r})\,r_a,\phi_{q}(\mathbf{r})\,\mathrm{d}\mathbf{r}, \\
 q_{pq} & = & - \sum_{a,b \in \{x,y,z\}} \lambda_a \lambda_b \int \phi^*_{p}(\mathbf{r})\,r_a r_b,\phi_{q}(\mathbf{r})\,\mathrm{d}\mathbf{r}.
\end{eqnarray}

As was mentioned above, only the left block carries the photonic operators, therefore  
the rules for building the action of the Hamiltonian on a trial wave function when employing the bipartite enlarged L-R splitting do not change along the DMRG sweep.
The purely photonic operators (third and fourth terms in Eq. \ref{ham_cs_fact}) can be added/absorbed into the left block Hamiltonian, which is coupled with an identity operator on the enlarged right block.

The second term in Eq. \ref{ham_cs_fact}) needs to be separated into several contributions with a different number of operators acting on the right block

\begin{eqnarray}
    \left\lbrace 
    -\sqrt{\frac{\omega}{2}}\sum_{\sigma}\sum_{p,q\in \mathrm{L}} d_{pq} a_{p\sigma}^\dagger a_{q\sigma} \left( b + b^\dagger \right)
\right\rbrace_\mathrm{L}&\otimes&
(\mathrm{id})_\mathrm{R}, \nonumber \\
    \sum_{\sigma}\sum_{q\in \mathrm{R}}
    \left\lbrace 
    -\sqrt{\frac{\omega}{2}}\sum_{p\in \mathrm{L}} d_{pq} a_{p\sigma}^\dagger \left( b + b^\dagger \right)
\right\rbrace_\mathrm{L}&\otimes&
(a_{q\sigma})_\mathrm{R}  - \mathrm{h.\,c.}, \nonumber \\
\sum_{\sigma}\sum_{p,q\in \mathrm{R}}
    \left\lbrace 
    -\sqrt{\frac{\omega}{2}}d_{pq} \left( b + b^\dagger \right)
\right\rbrace_\mathrm{L}&\otimes&
(a_{p\sigma}^\dagger a_{q\sigma}
)_\mathrm{R}.
\label{int_term}
\end{eqnarray}

\noindent
In our pilot implementation, we do not switch between the normal and presummed operators in the middle of each sweep, but perform presummations only on the left block. This way, we can absorb the first contribution from Eq. \ref{int_term} into the left block Hamiltonian. The left block part of the second contribution is then absorbed into $A_q^{\uparrow}$ and $A_q^{\downarrow}$ operators (\ref{opAup}, \ref{opAdown}), which are combined with a single annihilation operator on the enlarged right block. 
%Since the photonic operators act only on the very first site, this has to be performed only during the first blocking, i.e. when enlarging the photonic site by the first electronic site. 
Similarly, the left block part of the third contribution can be absorbed into the presummed quadratic operators, which are coupled with quadratic operators on the enlarged right block.

Another thing to note is that while the number of electrons needs to be kept fixed throughout the calculation, and forces us to keep track of the number of electrons and spin quantum numbers in each of the renormalized states, no such thing is necessary for the number of photons, as generally the wave function will contain contributions from states with different numbers of photons, and also the renormalized states will mix the number of photons together.  
This is in stark contrast to works combining bosonic degrees of freedom with quantum chemical DMRG for non-Born-Oppenheimer calculations\cite{NEO_reiher}, which preserve the number of bosonic particles. In solid state physics, however, the Hubbard Holstein model is used, which does not preserve the number of phonons \cite{DMRG_HH,DMRG_HH2,DMRG_HH3,block2}.

\section{Computational details}

\label{section_comp_details}
To demonstrate the performance of the cavity QED-DMRG method, we performed all-$\pi$ calculations of $n$-oligoacenes with $n$ ranging from 2 to 5, i.e. naphthalene, anthracene, tetracene, and pentacene. 
The geometries used were optimized (similarly to ref. \citen{DMRG-AC}) at the UB3LYP/6-31G(d,p) level. For all subsequent calculations we used the cc-pVDZ basis set.\cite{Dunning1989}
Naphthalene, which was recently analyzed using the QED-CASCI method \cite{QED-CASCI}, served as a basis for establishing the convergence of QED-DMRG with QED-CASCI results. For higher acenes, we investigate how the results, particularly the energy gap between polaritonic states, scale with the number of aromatic rings.

%The first thing to verify is the validity of the Cavity-QED DMRG method with respect to the original  
%Cavity-QED CASCI.\cite{QED-CASCI}
%For this we decided to use the naphtalene molecule, in the cc-pVDZ basis set with an active space containing all 10 $\pi$ orbitals. This active space is reasonably challenging for the DMRG, but small enough to be able to run reference CASCI to check the error in the energy. Then we decided to perform calculations also on higher acenes, to study the scaling with the number of rings.

The performance of QC-DMRG is well known to depend heavily on the type of orbitals used and their ordering on the 1D lattice \cite{amaya_2015}. To investigate the effect of orbital type on QED-DMRG, we compared the performance of two different sets of spatial orbitals. In addition to canonical orbitals, we employed Pipek-Mezey split-localized orbitals. The ordering of the orbitals was optimized using the Fiedler algorithm,\cite{fiedler:1973, fiedler:1975,PhysRevA.83.012508} based solely on the purely electronic part of the Hamiltonian. Due to our implementation, the photon site was fixed as the leftmost site in the chain.

Furthermore, we compared two different formulations of the PF-Hamiltonian presented in Section \ref{subsec_qed_dmrg}, namely the CS formulation (\ref{ham_cs}) and the PN formulation (\ref{ham_pf}). It is worth noting that in the case of QED-CASCI, the CS formulation was shown to perform better \cite{QED-CASCI}.
The two-site DMRG calculations were initialized using the CI-DEAS warm-up procedure \cite{LegezaCIDEAS} and employed either fixed bond dimensions or variable bond dimensions, achieving the predefined truncation error through the Dynamical Block State Selection (DBSS) approach \cite{legezaDBSS}.

As mentioned above, the comparison with QED-CASCI has been done on the naphtalene molecule, with the photon energy tuned to $\mathrm{S}_0$-$\mathrm{S}_1$ transition energy of $0.160984\;E_\mathrm{h}$. We used a coupling vector oriented in plane along the short axis of the molecule. The size of the coupling varied from 0.005 to 0.2 atomic units (a.u.).
%For studying the convergence with the number of photons a slightly exaggerated coupling vector value of $0.2$ a.u. was used.

\begin{figure}
    \centering
    \includegraphics[width=0.7\linewidth]{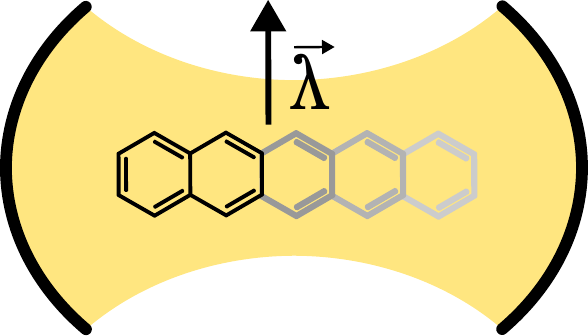}
    \caption{Orientation along the short axis of the oligoacene series, used throughout the majority of the calculations.}
    \label{fig:coupling orientation}
\end{figure}

%For the coherent localized formulation, we also tried using Dynamical Block State Selection aproach (DBSS)\cite{legezaDBSS} which varies the bond dimension to achieve a defined value of the the TRE.
%We used this to study the behavior of the error in energy close to convergence. 

Similarly for the larger acenes, where we only studied the behavior of the different states with the coupling strength, we used the photon energy equal to the $\mathrm{S}_0$-$\mathrm{S}_1$ transition at the zero coupling strength calculated with DMRG.
This gave us an excitation energy of $0.14083~E_\mathrm{h}$ for anthracene, $0.12741~E_\mathrm{h}$ for tetracene and  $0.11840~E_\mathrm{h}$ for pentacene.
We used a fixed bond dimension $M=1000$.
We compared two different orientations of the coupling vector, along the short axis of the molecule and along the long axis.

\section{Results and discussion}
\label{section_results}

\begin{figure}
    \centering
    \includegraphics[width=\linewidth]{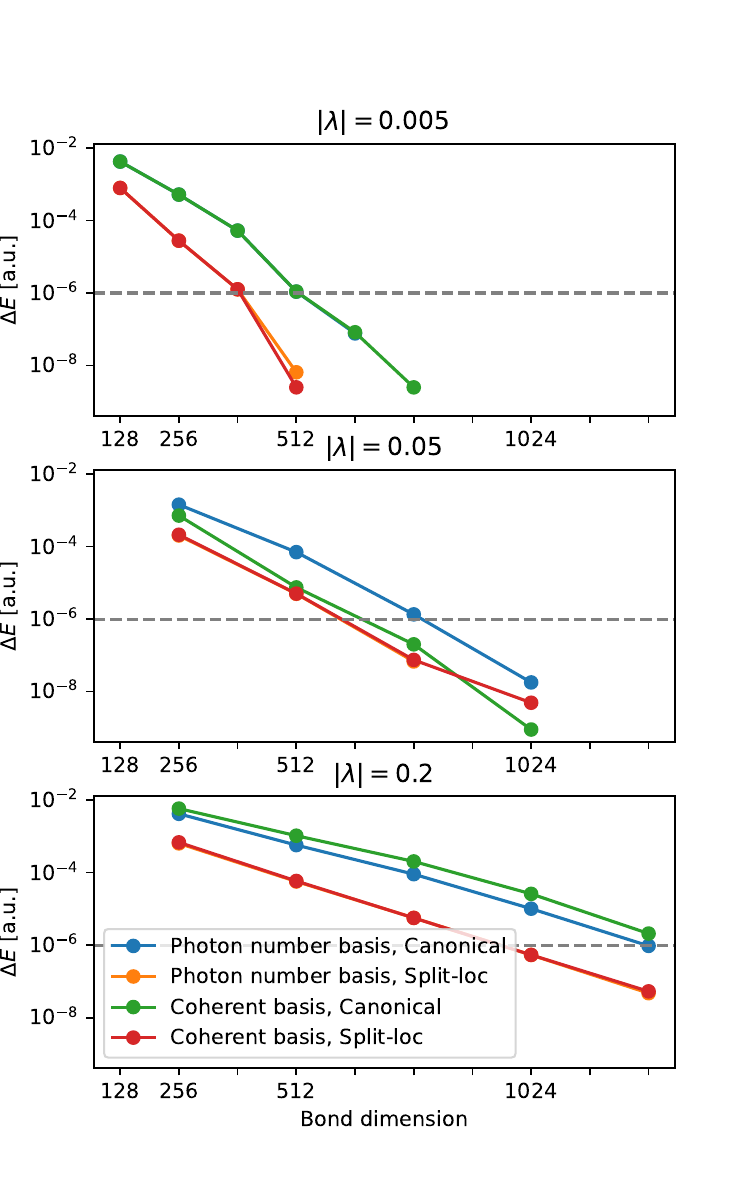}
    \vspace{-1.5cm}
    \caption{Comparison of the energetic errors in QED-DMRG with respect to exact QED-CASCI for the lower polaritonic state of naphthalene, using PN and CS Hamiltonian formulations in both canonical and split-localized MO bases. Results are shown for varying bond dimensions and three coupling strengths. The orange and red curves overlap in all three plots.}
    \label{fig:error}
\end{figure}

\begin{figure}
    \centering
    \includegraphics[width=\linewidth]{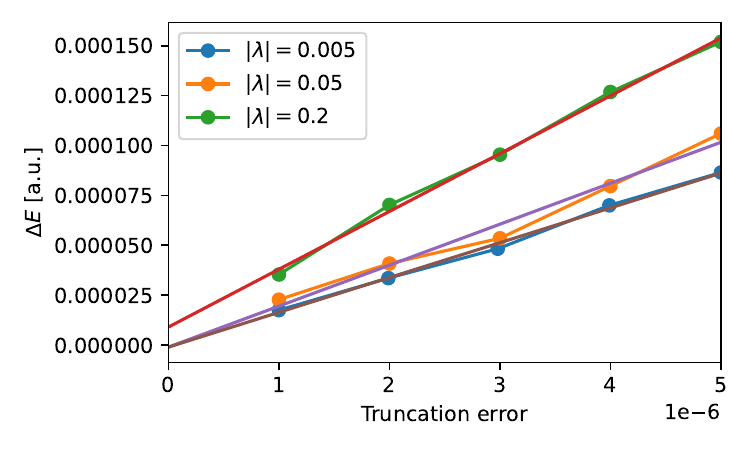}
    \caption{Dependence of the error in energy with the truncation error for the polaritonic state of naphthalene with different coupling strengths}
    \label{fig:extrapolation}
\end{figure}

\begin{table*}[t]
    \centering
    \caption{Truncation error of naphthalene with a bond dimension of 256 for different values of the coupling strength and the root mean square deviation (RMSD) of energies of the 5 lowest states from the exact CASCI energy.}
    \label{tab:TRE_scaling}
\begin{tabular}{llllllll}
\toprule
Lambda (a.u.)   & 0        & 0.005    & 0.01     & 0.02     & 0.03     & 0.04     & 0.05     \\
\midrule
Can-TRE  & 2.12E-03 & 2.12E-03 & 2.17E-03 & 2.21E-03 & 2.26E-03 & 2.35E-03 & 2.45E-03 \\
Loc-TRE  & 7.28E-05 & 1.49E-04 & 1.51E-04 & 1.52E-04 & 1.60E-04 & 1.68E-04 & 1.77E-04 \\\midrule
Can-RMSD & 5.71E-04 & 5.64E-04 & 5.63E-04 & 6.34E-04 & 6.87E-04 & 7.51E-04 & 8.21E-04 \\
Loc-RMSD & 2.60E-08 & 7.12E-05 & 7.37E-05 & 8.02E-05 & 9.10E-05 & 9.65E-05 & 1.04E-04\\
\bottomrule
\end{tabular}
\end{table*}

\begin{figure}
    \centering
    \includegraphics[width=\linewidth]{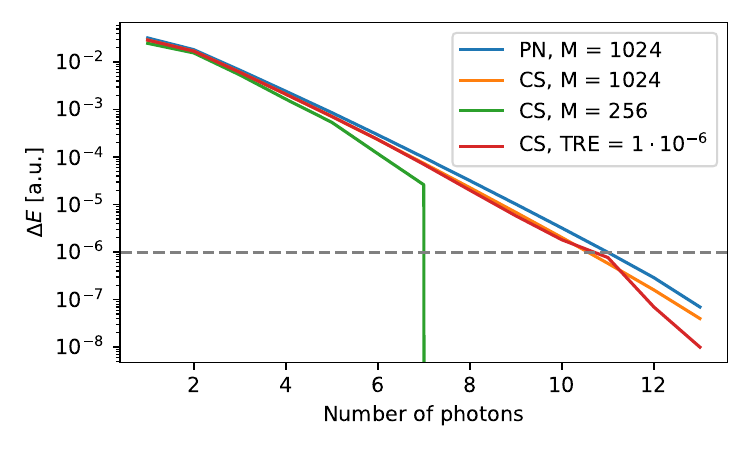}
    \caption{Energy difference between the polaritonic state with a given number of photons and the converged state with 14 photons, illustrating the saturation of the naphthalene wave function with photons. PN denotes the particle number basis, and CS denotes the coherent state basis.}
    \label{fig:ph_saturation}
\end{figure}

% \begin{table}[]
%     \centering
%     \caption{The maximum bond dimension for a given maximum number of photons reflecting the renomalization of unnecessary degrees of freedom. Naphtalene with a 0.2 coupling strength}
%     \label{tab:DBSS_saturation}
% \begin{tabular}{cc}
% \toprule
% Number of photons & Maximum bond dimension \\ \midrule
% 1   & 1247 \\
% 2   & 1418 \\
% 3   & 1490 \\
% 4   & 1524 \\
% 5   & 1543 \\
% 6   & 1552 \\
% 7   & 1556 \\
% 8   & 1557 \\
% 9   & 1558 \\
% 10  & 1558 \\
% 11  & 1558 \\
% 12  & 1558 \\
% 13  & 1558 \\
% 14  & 1558 \\
% \multicolumn{2}{c}{\vdots}
% \\
% 100               & 1558                   \\ \bottomrule
% \end{tabular}

% \end{table}

\begin{figure}
    \centering
    \includegraphics[width=\linewidth]{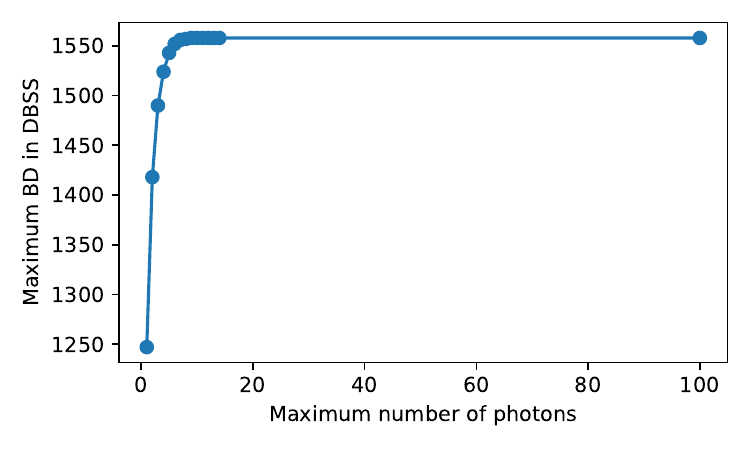}
    \caption{Maximum bond dimension required for a given maximum number of photons, illustrating the renormalization of unnecessary degrees of freedom. The analysis is based on naphthalene with a coupling strength of 0.2 and a target truncation error (TRE) of $10^{-6}$.}
    \label{fig:DBSS_saturation}
\end{figure}

\begin{figure}
    \centering
    \includegraphics[width=\linewidth]{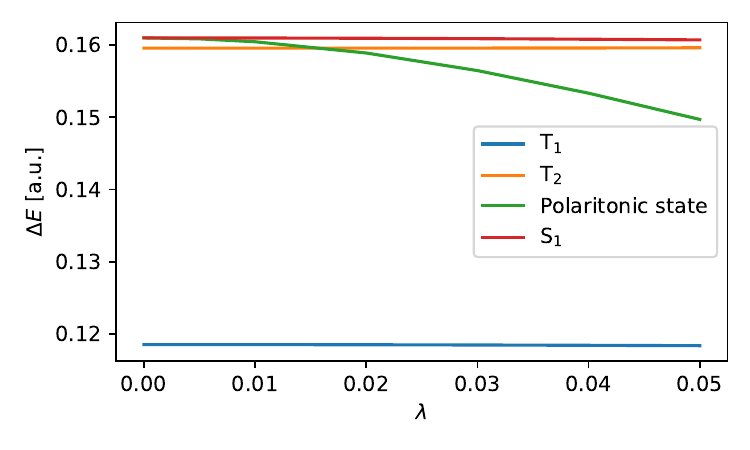}
    \caption{Energy differences between the ground state and excited states of the naphthalene molecule at a given coupling strength.}
    \label{fig:naphtalene_states}
\end{figure}

% \begin{figure}
%     \centering
%     \includegraphics[width=\linewidth]{Anthracene.pdf}
%     \caption{The anthracene molecule states energy difference from the ground state at a given coupling strength}
%     \label{fig:anthracene_states}
% \end{figure}

% \begin{figure}
%     \centering
%     \includegraphics[width=\linewidth]{Tetracene.pdf}
%     \caption{The 
%     tetracene molecule states energy difference from the ground state at a given coupling strength}
%     \label{fig:tetracene_states}
% \end{figure}

\begin{figure}
    \centering
    \includegraphics[width=\linewidth]{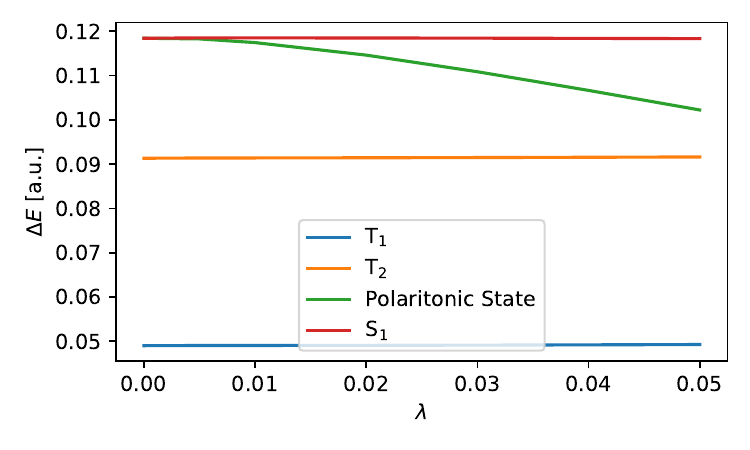}
    \caption{Energy differences between the ground state and excited states of the pentacene molecule at a given coupling strength.}
    \label{fig:pentacene_states}
\end{figure}

\begin{figure}
    \centering
    \includegraphics[width=\linewidth]{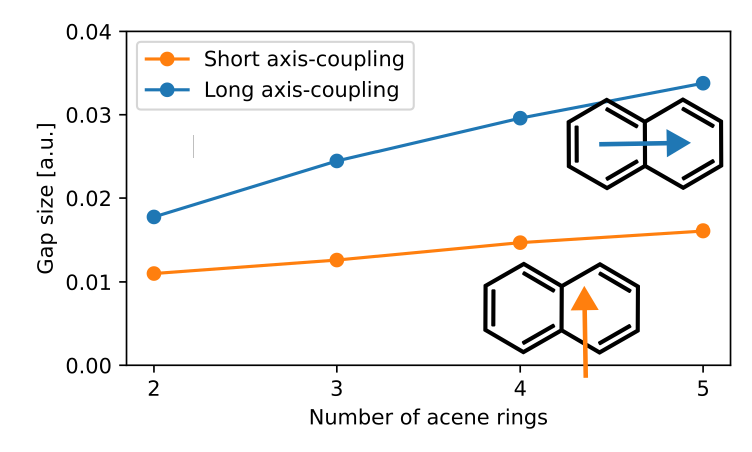}
    \caption{Energetic splitting of the polaritonic states for different  numbers of acene rings at a fixed coupling strength of $|\lambda| = 0.05$ a.u.. We compare two different possible orientations of the coupling vector.}
    \label{fig:gap_opening}
\end{figure}

Let's begin by comparing QED-DMRG and QED-CASCI on the naphthalene molecule.
Figure \ref{fig:error} illustrates the energetic error of QED-DMRG relative to QED-CASCI for the lower polaritonic state across three different coupling strengths. Increasing the bond dimension allows us to reduce the error below our convergence threshold of $10^{-6}$ $E_{\text{h}}$ for DMRG.
The almost linear dependence in a semi-logarithmic plot hints at a roughly exponential decrease of the error with an increasing bond dimension \cite{DMRG_exponential, chan2002highly,chan_review}. 
%Other, more sophisticated estimates are known, however they are in accord with this roughly exponential convergence.\cite{chan2002highly,chan_review}
One can observe that with increasing coupling strength, the error for a given bond dimension grows due to greater entanglement between photonic and electronic degrees of freedom.

Similarly to QC-DMRG, we can see in Figure \ref{fig:error} that the truncation error is strongly reduced by localization of the active orbitals, the truncation error being an order of magnitude lower with the same bond dimension. Thus, for calculations of higher acenes, we have employed the split-localized orbitals. Overall, we found that for naphthalene with a split-localized basis we can achieve submiliHartree accuracy with respect to QED-CASCI quite easily, even with a rather small bond dimension of 256.

In Table \ref{tab:TRE_scaling} we show the truncation error and the RMSD of the five lowest states with respect to the reference CASCI depending on the magnitude of the cavity coupling. Similarly as with the error in energy, the truncation error grows with increasing the coupling strength. However, the growth is not very dramatic, with less than a $20\%$ increase of the TRE with an order of magnitude stronger coupling.

Figure \ref{fig:extrapolation} shows the dependence of the energy error on the TRE set for various coupling strengths. A linear trend is observed for small TRE values, similar to QC-DMRG, which enables extrapolation of the energy to a zero TRE limit, which should in theory be equivalent to the exact CASCI energy.  The errors of the extrapolated energies are $-1 \cdot 10^{-6}$, $-1 \cdot 10^{-6}$, and $9 \cdot 10^{-6}$ atomic units for coupling values of $|\lambda| = 0.005$ a.u., $|\lambda| = 0.05$ a.u., and $|\lambda| = 0.2$ a.u., respectively. These errors are significantly lower than those of the most accurate calculations used for the extrapolation, which had a TRE of $1 \cdot 10^{-6}$. 

In Figure \ref{fig:ph_saturation} we show the error in energy of the polaritonic state of naphthalene with respect to the maximum number of photons in the wave function. The error is calculated for the coupling strength $\lambda=0.2$ a.u. as a difference between the energy with a given number of photons and the energy with 14 photons, which we considered converged.
We can observe that the error decreases roughly exponentially with the number of photons, similar to ref. \citen{QED-CASCI}. Nevertheless, the approximate nature of DMRG introduces additional features worth discussing. Using a bond dimension of 1024, which yields energies almost identical to the reference CASCI, we observe proper scaling across the entire range of values. Conversely, reducing the bond dimension to 256 limits the flexibility of the wave function, making it unable to fully account for the number of photons.
As such, the system artificially gets saturated with a much smaller number of photons.
Increasing the number of photons beyond this saturation limit results in nearly no changes to the resulting wave function and the change in energy abruptly drops to zero.
Similarly, with DBSS the resulting curve follows the scaling until the energy error from the photon truncation does not reach roughly $10^{-6}$, which would be roughly the expected accuracy of a calculation with a fixed truncation error of $10^{-6}$. The convergence does not differ very significantly between the photon number and coherent basis formulations, however, the coherent formulation has the advantage of being translationally invariant, which in the photon number basis does not hold for charged molecules, unless we include enough photons in the calculation to saturate the wave function.\cite{Foley_CPR_2023,DePrince23_5264,QED-CASCI} On the other hand, the coherent state formulation uses the QED-RHF orbitals, which change with the coupling strength, and requires selecting an active space for every coupling vector separately, unlike with canonical RHF orbitals.

One advantage of the DMRG method we would like to stress is that the extra photons in the basis, which do not contribute to the wave function, are removed in the renormalization procedure, and thus mean a very small extra computational cost. This is in sharp contrast with QED-CASCI, where every extra photon inevitably means an increase in the Hamiltonian matrix size.
To demonstrate this, in Figure \ref{fig:DBSS_saturation} we show the maximal bond dimension for DBSS for each number of photons. 
There we can clearly see that increasing the number of photons after saturating the wave function has no effect on the total number of renormalized basis states, even when going to the extreme limit of allowing up to 100 photons.

To explicitly show the polaritonic splitting of the S$_1$ excited state of naphthalene, in Figure \ref{fig:naphtalene_states} we plot the dependence of excitation energies on the coupling strength.
We find that, except for the split polaritonic state, the other states are affected by the interaction with the field equally, and thus the energy difference stays constant over the range.
The split polaritonic state is then lowered in energy with respect to the other states, which can lead to states crossing, as can be seen for the $\mathrm{S_1}$ and $\mathrm{T_2}$ states in this case. The primary determinantal contributions to the resulting polaritonic state of naphthalene for $\lambda = 0.05$ a.u., with absolute CI coefficients greater than 0.01, are

\begin{eqnarray}
    \ket{\Psi} & = & -0.883 \left|\text{HF det.}\right\rangle \left| 1 \omega\right\rangle + 
    0.119 \left|H-1\rightarrow L\right\rangle \left| 0 \omega\right\rangle \nonumber \\
     & + & 0.119 \left|\overline{H-1}\rightarrow \overline{L}\right\rangle \left| 0 \omega\right\rangle +
    -0.116 \left|H\rightarrow L+1\right\rangle \left| 0 \omega\right\rangle \nonumber \\
     & + & -0.116 \left|\overline{H}\rightarrow \overline{L+1}\right\rangle \left| 0 \omega\right\rangle +
     0.105 \left|H,\overline{H}\rightarrow L,\overline{L}\right\rangle \left| 1 \omega\right\rangle \nonumber, \\
\end{eqnarray}

\noindent
where $H$ and $L$ denote the highest occupied molecular orbital (HOMO) and the lowest unoccupied molecular orbital (LUMO).

Similarly, we show the value of the polaritonic splitting of pentacene in Figure \ref{fig:pentacene_states}. 
Although the polaritonic splitting increases with the number of aromatic rings, due to the larger 
$\mathrm{S}_1 - \mathrm{T}_2$
gap, the state inversion occurs at higher coupling strengths. 

By plotting the splitting value for a fixed coupling strength of 
$|\lambda| = 0.05$ a.u. in Figure \ref{fig:gap_opening}, we observe a remarkable linear increase with the number of rings, despite the underlying complexity of the processes involved. This trend, which is more pronounced for the coupling oriented in plane along the long axis, highlights the influence of molecular structure on polaritonic behavior and suggests potential avenues for tuning the properties of polycyclic aromatic hydrocarbons through molecular design.

\section{Conclusions}
\label{section_conclusions}

In conclusion, we have developed a novel method for cavity-QED calculations utilizing the DMRG algorithm, which facilitates near exact CASCI calculations with significantly larger active spaces than traditional canonical-CASCI method. This approach is particularly suitable for systems exhibiting strong correlation effects between electronic and photonic degrees of freedom. We have demonstrated the method's capabilities on the $n$-oligoacenes series, with $n$ ranging from 2 to 5, successfully managing up to 22 fully correlated $\pi$ orbitals in pentacene. Our numerical results indicate that employing a split-localized coherent state basis yields fastest convergence towards the exact results. Additionally, we have shown that in the acenes series, the polaritonic splitting increases almost linearly with the number of aromatic rings, highlighting the influence of molecular structure on polaritonic behaviour.

\section*{Acknowledgment}
We acknowledge financial support from the Czech Science Foundation (grant no. 23-05486S), the Charles University Grant Agency (Grant No. 218222), the Ministry of
Education, Youth and Sports of the Czech Republic through the e-INFRA CZ (ID:90254), the Advanced Multiscale Materials for Key Enabling Technologies project, supported by the Ministry of Education, Youth, and Sports of the Czech Republic. Project No. CZ.02.01.01/00/22\_008/0004558, Co-funded by the European Union. JJF, NV, and NG acknowledge support from the Center for Many-Body Methods, Spectroscopies, and Dynamics for Molecular Polaritonic Systems (MAPOL) under FWP 79715, which is funded as part of the Computational Chemical Sciences (CCS) program by the U.S. Department of Energy, Office of Science, Office of Basic Energy Sciences, Division of Chemical Sciences, Geosciences and Biosciences at Pacific Northwest National Laboratory (PNNL). PNNL is a multi-program national laboratory operated by Battelle Memorial Institute for the United States Department of Energy under DOE contract number DE-AC05-76RL1830.  JJF gratefully acknowledges the NSF CAREER Award CHE-2043215.
\section*{References}
\bibliography{references,foley_polariton_refs}   

\end{document}

%% file: introduction.tex
Photonic cavities are an excellent tool for studying the interactions between the quantised electromagnetic field and matter. Originally starting with Rydberg atoms\cite{CQED_1992,CQED_1993}, experimental developments enabled the field to move to larger systems, such as organic chromaphores\cite{meth_blue}.
This opened a myriad of new potential applications, including polaritonic chemistry, where chemical reactions are modified through strong coupling between cavity modes and molecular electronic or vibrational degrees of freedom~\cite{Polchem1,Polchem2,Polchem3,polchem4}. %which uses the ability to modify the landscape for chemical reactions using strong coupling to an optical cavity
These developments go hand in hand with the need for accurate theoretical methods, allowing us to be confident in both our predictive abilities and the interpretation of experimental results~\cite{fabijan_click}.

A detailed understanding of molecular structure and dynamics when cavity modes strongly couple to molecular electronic degrees of freedom (i.e. electronic strong coupling) requires that both electronic and photonic degrees of freedom are treated on equal quantum mechanical footing.  One route that has been pursued by several groups recently includes generalizing the tools of {\it ab initio} electronic structure theory to explicitly include coupling to quantized photonic degrees of freedom. Such approaches have included quantum electrodynamics generalizations of density functional theory (QEDFT~\cite{Bauer11_042107,Rubio14_012508,Rubio15_093001,Rubio18_992,Appel19_225,Narang20_094116,Rubio22_ChemRev} and QED-DFT\cite{DePrince22_9303,Rubio22_094101,DePrince23_5264}),
real-time~\cite{Bauer11_042107,Tokatly13_233001,Rubio14_012508,Rubio17_3026,Tokatly18_235123,Varga22_194106} and linear-response~\cite{Shao21_064107,Shao22_124104,DePrince22_9303} formulations of QED-TDDFT, configuration interaction (QED-CIS)~\cite{Foley_154103, QED-CASCI}, cavity QED extension of second-order M{\o}ller-Plesset perturbation theory and the algebraic diagrammatic construction~\cite{Bauer_JCP_2023, Reichman_JCTC_2024}, coupled cluster (QED-CC)~\cite{Koch21_094113, Manby20_023262,DePrince21_094112,DePrince23_5264}, variational QED-2-RDM methods~\cite{DePrince_PRA_2022}, and diffusion quantum Monte Carlo (QMC).~\cite{Zhang_PRA_2024}.  Many of the aforementioned methods, however, cannot properly describe strong (multireference) electronic correlation present in many interesting molecular systems, such as transition metal complexes.  The recently-reported QED-CASCI approach\cite{QED-CASCI}
and QED-2-RDM methods are well suited for describing strong correlation, but both have 
important limitations. QED-2-RDM methods have polynomial scaling with respect to the active space size, but so far, these variational approaches are designed to simulate the lowest energy states of a given spin symmetry, which are of limited utility for studying polariton states~\cite{DePrince_PRA_2022}.  The QED-CASCI method can be used to compute multiple states, but scales exponentially with the active space size, putting a hard limit of less than $~20$ correlated orbitals~\cite{QED-CASCI}.  The exponential scaling of CASCI is usually surpassed by using approximations to  such as Selected-CI \cite{asci, Garniron2018, Liu2016}, heat-bath CI \cite{Sharma2017}, full-CI QMC \cite{neci}, or DMRG \cite{chan_review, Szalay2015, reiher_perspective}.  To this end, we present a QED extension to the Density Matrix Renormalization Group method (QED-DMRG) that can provide efficient controlled approximations to numerically-exact solutions
to the Pauli-Fierz Hamiltonian\cite{Rubio22_ChemRev} 
%it is not polynomial in general case
%there is no classical method with polynomial scaling...
%with polynomial cost.  
QED-DMRG also provides a powerful representation for approximating multiconfigurational multicomponent wavefunctions with similar accuracy and significantly lower computational cost than the recently-reported QED-CASCI approach.